**Anomalous spin-pumping behavior of half-metallic ferromagnet/d-wave superconductor heterostructures**

Hadi H. Hassan,[1] Santiago J. Carreira[1,*], M. Cabero[2], F. Martinet[1], Alexander Buzdin[3], Jacobo Santamaria[4], and Javier E. Villegas[1,+]

[1]*Laboratoire Albert Fert, CNRS, Thales, Université Paris Saclay, 91767, Palaiseau, France*
[2]*ICTS - Centro Nacional de Microscopía Electrónica, Universidad Complutense de Madrid, 28040 Madrid, Spain*
[3]*Condensed Matter Theory Group, University Bordeaux I, CPMOH UMR-CNRS 5798, 33405 Talence Cedex, France*
[4]*GFMC, Dept. Física de Materiales, Facultad de Física, Universidad Complutense, 28040 Madrid, Spain*

Spin-pumping experiments in superconductor/ferromagnet heterostructures, which probe spin-sinking by the superconductor, have revealed a variety of complex behaviors. Most experimental studies have focused on conventional s-wave superconductors combined with metallic or insulating ferromagnets. Here, we study a d-wave superconductor paired with a half-metallic ferromagnet, in epitaxial $YBa_2Cu_3O_{7-\delta}/La_{0.7}Sr_{0.3}MnO_3$ heterostructures with two crystalline orientations: one in which YBCO is c-axis oriented, and the other in which YBCO grows along the (103) direction. Using ferromagnetic resonance (FMR), we probe the temperature-dependent Gilbert damping coefficient α. For (103) heterostructures, α(T) initially decreases below Tc, but then increases at lower temperatures, eventually exceeding normal-state levels. This behavior can be understood in terms of the opening of the superconducting gap and spin transport via nodal quasiparticles, whose contribution is enhanced when the ab-plane of YBCO is exposed at the interface. In stark contrast, c-axis heterostructures exhibit a pronounced enhancement of α(T) below $T_C$, peaking at ~0.65–0.7$T_C$ before decaying. This anomaly suggests the dominance of interfacial Andreev bound states, arising from a locally suppressed superconducting order parameter due to proximity effects with the half-metallic LSMO.

* santiago.carreira@cnrs-thales.fr
+ javier.villegas@cnrs-thales.fr

## I. INTRODUCTION

The interplay between the antagonistic superconductivity and magnetism has been studied by physicists for years, using as a playground hybrid superconductor/ferromagnet heterostructures and devices [1–4]. Among the plethora of explored effects, which includes for example, superconducting proximity [5,6], magnetostatic coupling [7,8] or spin injection [9–12], the coupling between the superconducting state and the magnetization dynamics in various forms has become a pivotal topic [13–19], both for its fundamental interest and for its relevance in the nascent field of superconducting spintronics [20].

One of the key tools for investigating the dynamic coupling between superconductivity and magnetism is ferromagnetic resonance (FMR). In FMR experiments, a dc field $H$ is applied to saturate the macroscopic magnetization $\vec{M}$ and a microvawe field $h_{rf}$ [see Fig. 1] is applied perpendicular to $H$, driving a magnetization precession around the equilibrium axis defined by the dc field. The dynamics of the magnetization is described by the Landau-Lifshitz-Gilbert (LLG) equation [21]:

$$\frac{\partial\vec{M}}{\partial t}=\gamma\vec{M}\times\vec{H}_{eff}-\frac{\alpha}{M_s}\vec{M}\times\frac{\partial\vec{M}}{\partial t} \quad (1)$$

where $\vec{H}_{eff}$ is the effective field, $M_s$ is the saturation magnetization, $\gamma$ the gyromagnetic ratio, and $\alpha$ the Gilbert damping coefficient –related to the magnetization's relaxation rate of towards equilibrium. Typically, the experiments are carried out by fixing the microwave field frequency $f$ and sweeping $H$ across the resonance field $H_r$, which is related to $f$ by the Kittel formula [22]

$$f=\mu_0\frac{\gamma}{2\pi}\sqrt{(H_r+H_k)(H_r+H_k+M_{eff})} \quad (2)$$

where $H_k$ is the anisotropy field, and $M_{eff}$ is the effective magnetization. This resonance, which yields a measurable enhancement of the microwave absorption, presents a characteristic linewidth resulting from frequency-independent broadening $\Delta H_0$ (related to magnetic inhomogeneities) and a frequency-dependent one that is proportional to the damping coefficient $\alpha$ [21]. This reflects both intrinsic relaxation mechanisms (into the ferromagnet's crystal lattice) and extrinsic ones. Particularly, in the presence of an adjacent non-magnetic overlayer, $\alpha$ is sensitive to the spin-pumping phenomenon [23]: the precession of magnetization creates a non-equilibrium interfacial spin accumulation that diffuses into the non-magnetic layer, yielding a pure spin current $j_S$ (Fig. 1b). The angular momentum leakage into the adjacent layer, which therefore behaves as a "spin sink", leads to an enhancement of $\alpha$. Consequently, FMR is an ideal tool for studying spin dynamics in superconducting (S) materials [24]. On the one hand, it allows the creation of a spin current in the S material via spin-pumping. On the other hand, studying $\alpha(T)$ allows detecting how the S transition state changes the spin current in the superconductor.

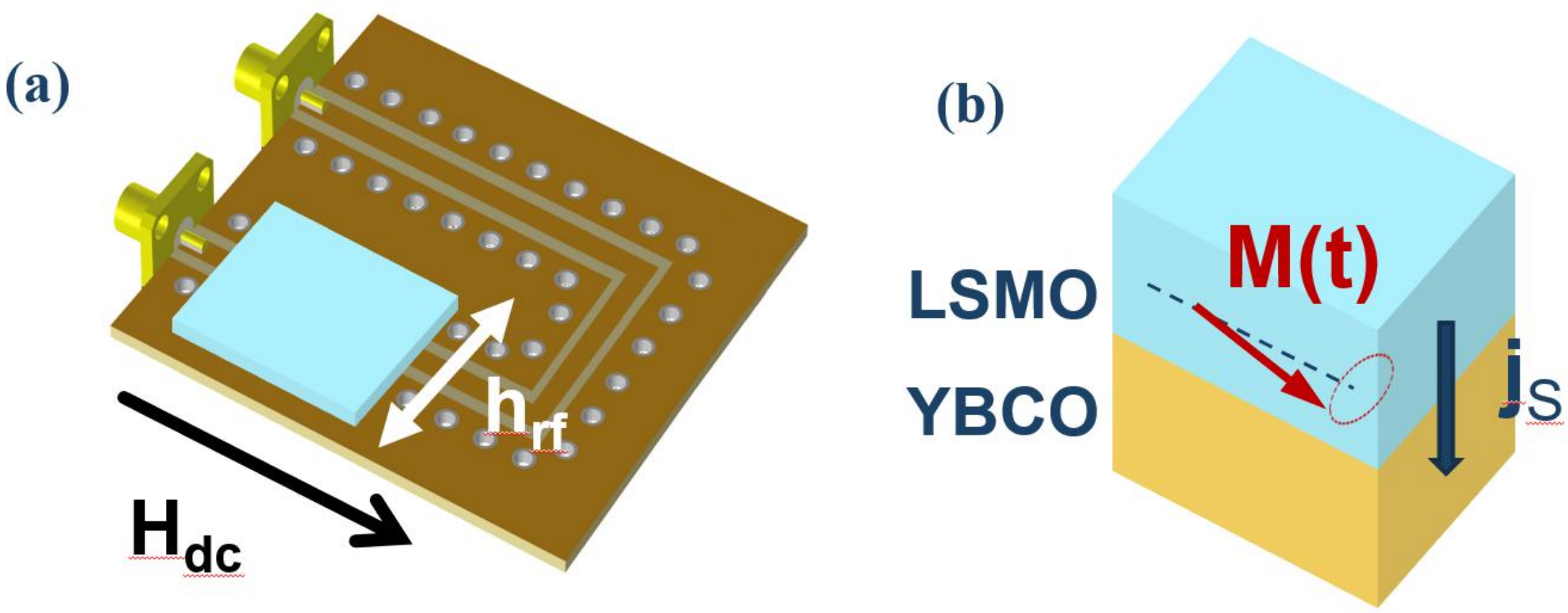

**Fig 1 :** (a) sketch of the coplanar waveguide and the sample mounted on top, where the DC and rf fields directions are indicated. (b) sketch of the LSMO/YBCO bilayer, where the processing magnetization in the LSMO yield to spin pumping and a spin current into YBCO**.**

The pioneering FMR experiments in superconductor/ferromagnet heterostructures were conducted on bilayers made of the metallic permalloy (Py) and the conventional s-wave superconductor Nb [18]. These experiments, similarly to more recent ones on $MgB_2$ [25], found a drastic shrinking of the FMR linewidth across the superconducting transition. This was explained by considering the opening of the S gap, with the consequent vanishing density of electron states at the Fermi level, which hinders the spin flow into the S material. Such a blocking effect would strengthen as the temperature decreases because the concentration of thermally excited low-energy quasiparticles (QPs) diminishes [26]. In contrast, another work based on insulating F/s-wave S (GdN/NbN) found a completely different behavior: in this system, $\alpha(T)$ showed a peak below the S transition [16] and subsequently dropped below the normal-state level upon further temperature decrease. This novel behavior was initially ascribed to spin-orbit effects [16]. Later theoretical work [27] proposed that it could be explained considering the so-called coherence peaks in the QP density of states. These appear at the gap edges and come close to the Fermi level in the vicinity of $T_C$, providing a high density of states to absorb and carry spin into the superconductor, yielding a peak in $\alpha(T)$ right below $T_C$ [24]. The role of these gap-edge QPs in transporting spin into the superconductors was also discussed to explain anomalously high inverse spin Hall signals in the vicinity of $T_C$ of superconductor/ferromagnet devices [28]. More recent theoretical work proposed that the drastic enhancement of spin pumping right below $T_C$ may also be explained if superconductivity is depressed near the interface, that is, with a local $T_C' < T_C$ in the bulk of the superconductor [29]. For T near $T_C$, this would result in QP interface-bound states that significantly contribute to enhancing spin transport into the superconductor. [29] At variance to all the work discussed above, experiments by Jeon *et al.* [17] on Py/Nb/Pt heterostructures found a gradual, steady spin pumping enhancement upon cooling below $T_C$. This behavior was ascribed to spin transport by unconventional equal-spin triplet superconducting pairs, whose emergence would be favored by the spin-orbit coupling in Pt [30], possibly in combination with exchange field [31] and Landau Fermi liquid [32] effects. In summary, the variety of behaviors observed in all those spin pumping experiments has been explained based on two main possibilities for spin transport into superconductors: quasiparticles [18,27,29] or equal-spin triplet pairs [30].

While the work summarized above focused on conventional s-wave superconductors, experiments based on the d-wave superconductor $YBa_2Cu_3O_{7-d}$ (YBCO) and metallic ferromagnets Py were reported by some of us [33]. In this system, the behavior varied significantly with interfacial properties. The results obtained for heterostructures with flat interfaces oriented along the crystallographic c-axis (001) of YBCO were similar to those of s-wave/Py ones [18], with a drastic reduction of spin pumping right below $T_C$ and a subsequent stabilization well below the normal-state level, explained by the opening of the superconducting gap. However, in the presence of rough interfaces that partially expose the *ab* plane to the ferromagnet, the drop at $T_C$ was followed by a steady increase in spin pumping, eventually exceeding the normal-state level upon further temperature decrease. This unusual low-temperature behavior was explained considering the presence of nodal QPs associated with the d-wave pairing [33]. Shortly after, theoretical studies on d-wave superconductor/ferromagnetic insulator (FI) interfaces [34] confirmed the role of nodal QPs in enhancing spin pumping at low temperatures –in this case, only to a level that remains below the normal-state one.

In the present work, we investigate spin pumping in heterostructures composed of the half-metallic ferromagnetic manganite $La_{0.7}Sr_{0.3}MnO_3$ (LSMO) and the $d$-wave cuprate superconductor YBCO. The interfacial magnetism [35–39], superconducting proximity effects [40–49], and spin-dependent transport of the family of cuprate/manganite heterostructures [47,50–52] have been widely investigated. One of the salient findings is the evidence for proximity-induced equal-spin triplet superconductivity in c-axis YBCO/LSMO heterostructures [45,53,54] (as well as in the similar YBCO/$La_{0.7}Ca_{0.3}MnO_3$ ones [40–44]). Here, we conduct spin-pumping experiments in c-axis YBCO/LSMO bilayers and carry out a comparison with heterostructures grown with a different crystalline orientation, namely (103) YBCO/ (110) LSMO. The latter presents a spin-pumping reduction across the superconducting transition, followed by an increase above the normal-state level upon further temperature decrease. This is expected considering earlier experiments [33] because, with this crystalline orientation, the YBCO's *ab* plane is largely exposed at the interface, favoring the role of nodal QPs. In contrast, c-axis heterostructures show a drastic enhancement of spin pumping below the superconducting transition, which is opposite to the behavior observed earlier in c-axis YBCO/Py heterostructures with smooth interfaces [33]. Considering the superconducting proximity effect characteristic of c-axis YBCO/manganite interfaces [40–45,53,54], and the strong similarity between our experimental results and theoretical calculations of spin pumping across S/FI interfaces with locally depressed superconductivity, we ascribe the unusual spin-pumping behavior of these heterostructures to the presence of interfacial Andreev bound states [29]. Overall, our results point to the coexistence of different mechanisms for spin transport across the d-wave/half-metal interface, whose dominance depends on the crystalline orientation because this determines the access to nodal QPs and the strength of proximity effects.

## II. SAMPLE FABRICATION AND STRUCTURAL CHARACTERIZATION.

Epitaxial YBCO (top) / LSMO (bottom) bilayers were grown on $NdGaO_3$ (NGO) substrates. The choice of NGO as substrate is motivated by the good structural matching with YBCO and LSMO, as well as by its low microwave absorption at low temperatures [55] –which is relevant for the present FMR experiments. We chose substrates with two different orientations, (110) and (100). As demonstrated below, this allows obtaining YBCO/LSMO interfaces with different crystalline orientations. Growth on (110) NGO yields (001) YBCO/(001) LSMO, that is, a c-axis orientation, which is the most commonly studied in the literature [39–41,45,50,53,54,56–58]. Growth on (100) results in (103) YBCO/ (110) LSMO heterostructures –hereafter called (103) samples. While (103) YBCO single films have been grown and studied earlier in different contexts [59–61], to the best of our knowledge, epitaxial (103) YBCO/ (110) LSMO heterostructures have not been explored earlier. The growth was done by Pulsed Laser Deposition (PLD) using a KrF excimer laser. The laser fluence was 1 J/cm$^2$ for LSMO and 1.5 J/cm$^2$ for YBCO. During deposition, the substrate temperature was 700 ºC and the oxygen pressure 0.36 mbar. After deposition, samples were cooled down to room temperature in 1 bar of pure oxygen. In what follows, we compare results from eight different samples, all of them with lateral dimensions ~ 5 mm x 5 mm. Four of these samples are superconducting: one (103) sample with YBCO and LSMO 20 nm and 50 nm thick, respectively, and three c-axis heterostructures grown on NGO (110), with a fixed YBCO thickness (20 nm) and varying LSMO thicknesses –47 nm, 17 nm and 13.8 nm. For each of those, a control counterpart was grown having the same LSMO thickness but much thinner (5 nm) YBCO, which results in the suppression of superconductivity. Comparison of each superconducting (S) heterostructure with its non-superconducting (nS) counterpart allowed us to identify superconducting effects in the spin-pumping behavior.

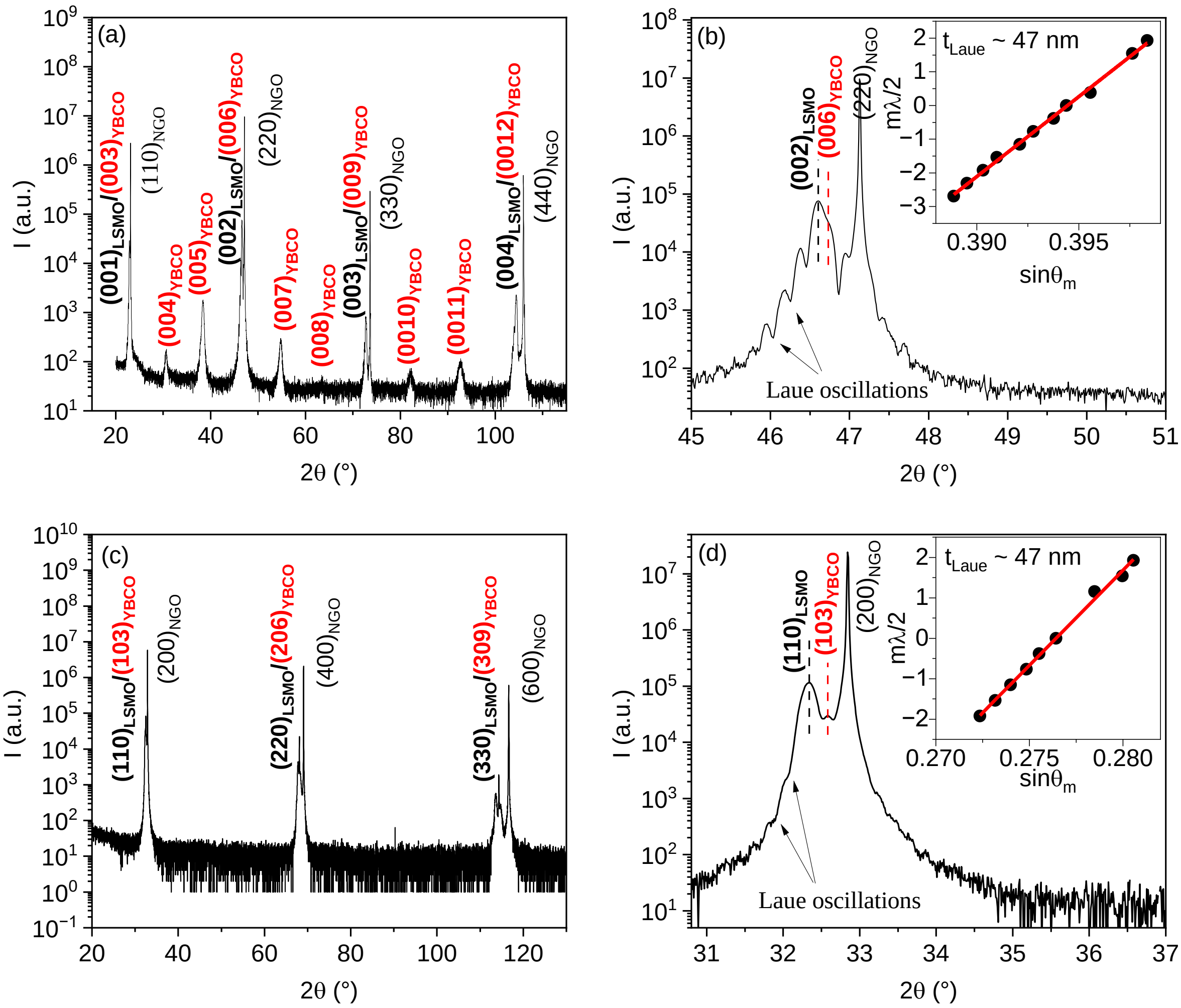

**FIG. 2**. (a-c) X-ray diffraction (XRD) of a LSMO-YBCO bilayer grown on (a) NGO (110) and (c) NGO (100). In (b) and (d) we show in detail the Bragg peaks of LSMO and YBCO around the (b) NGO (220) and (d) NGO (200). The linear trends in the insets in (b) and (d) were calculated from the maxima and minima of the Laue oscillations around the LSMO peak. The XRD measurement shows epitaxial growth of the bilayer along the YBCO c-axis, while the calculated Laue thicknesses are in agreement with the nominal LSMO thickness.

The structural characterization of the heterostructures included X-ray diffraction (XRD) and high-resolution Scanning Transmission Electron Microscopy (STEM). $\theta$-$2\theta$ XDR scans for samples grown on (110) NGO [see example in Fig. 2 (a)] only show (00n) Bragg reflections for both YBCO and LSMO, demonstrating epitaxial growth along the c-axis. A closer look at LSMO's Bragg peaks (Fig. 2 b) reveals finite-size Laue oscillations, which further evidence a good structural quality, and indicate that layer thickness is homogeneous at the unit-cell scale.

The thickness can be calculated from the oscillations' period by indexing the maxima and minima around the Bragg peak. If the fringes are indexed by an integer $m$, the $\theta_m$ positions of the maxima and minima with respect to the Bragg peak $\theta_B$ is described by the expression [62],

$$sin(\theta_m)\text{-}sin(\theta_B)=\frac{m\lambda}{2t} \qquad (3)$$

where $\lambda$ is the X-ray wavelength and $t$ is the film thickness. Examples of linear fits with the calculated thickness are shown in the inset of Fig. 2b and Fig. 2d.

Samples grown on (100) NGO only show (n n 0) reflections for LSMO and (n 0 3n) ones for YBCO (Fig. 2c and 2d), demonstrating epitaxial growth of both layers. Laue oscillations are in this case weaker, suggesting a slightly rougher interface. It is noteworthy that the (103) and (110) Bragg reflections of YBCO lie at similar angles in a scan, so the distinction of these phases through specular XRD experiments is not straightforward. However, as we shall see next, STEM images confirm the YBCO (103) phase.

Atomic-resolution HAADF-STEM images were acquired on a JEOL ARM200cF operated at 200 kV. Figure 3 (a) shows a cross-sectional view of a c-axis YBCO/LSMO bilayer, revealing an atomically abrupt interface that confirms the high crystalline quality of the films. Occasionally, a local change in the atomic stacking sequence can be observed at the interface, which gives rise to a planar defect extending vertically through the entire YBCO layer, as shown by the yellow-boxed region of Figure 3 (a). This yields an antiphase boundary (APB), marked by a half-unit-cell displacement (~0.2 nm) that disrupts the regular perovskite cation ordering [63,64]. Figure 3 (b) provides a magnified view of the APB. The intensity line profiles below –extracted across the regions between red and white arrows– compare the HAADF

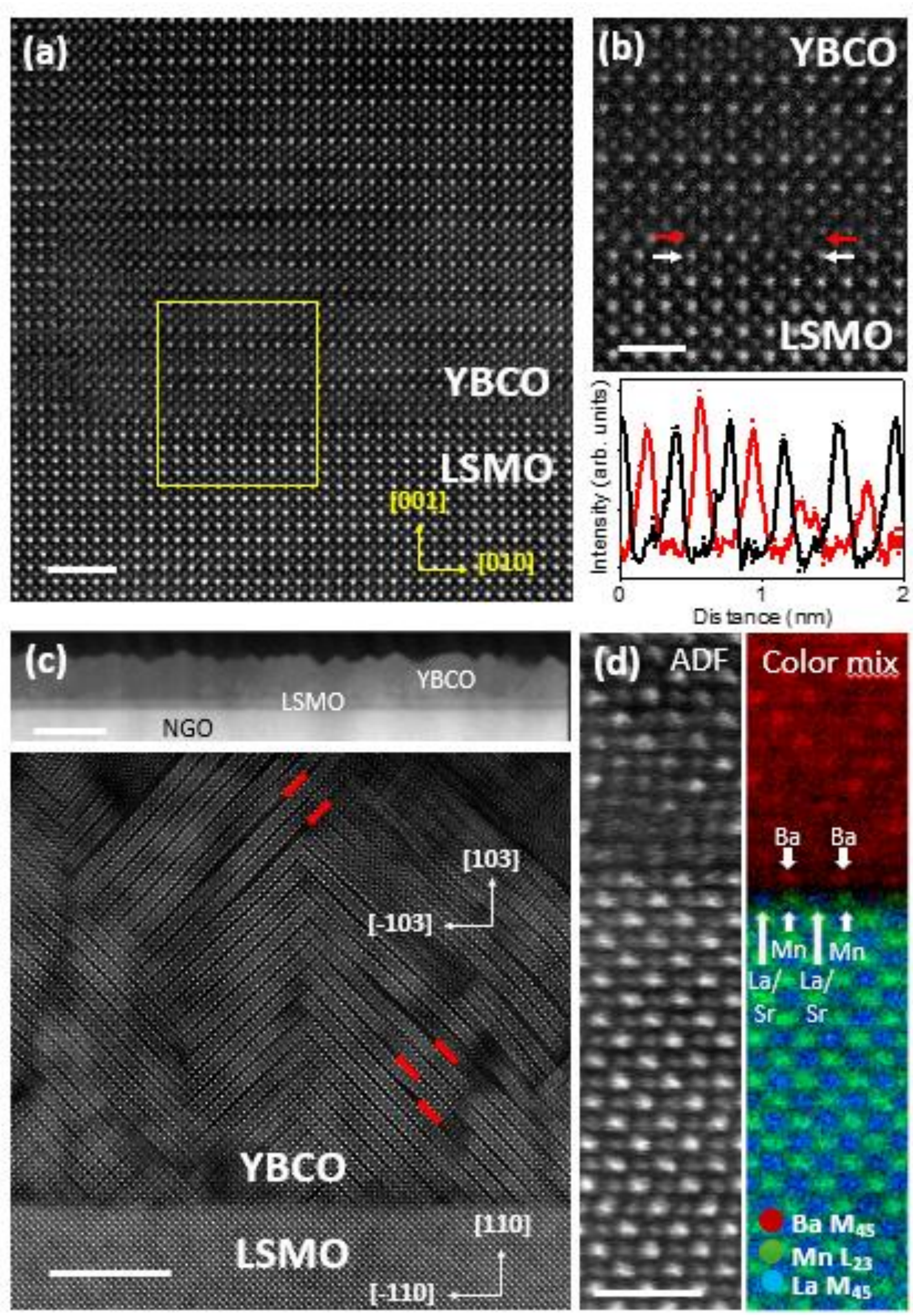

**Figure 3** (a) Cross sectional HAADF-STEM image of an LSMO/YBCO interface. (b) Higher magnification view of the yellow boxed region in (a), with the red/white arrows indicating the red/black intensity profiles of the cationic columns shown below (red for APB plane, black for the one below it). **Scale bars:** (a) 2 nm and (b) 1 nm. (c) Cross sectional HAADF-STEM images of an LSMO/YBCO sample grown on a NGO [110] substrate. Upper panel reveals the substrate-film interfaces and surface roughness of the YBCO layer. The lower panel displays an atomic resolution image of the atomically sharp LSMO/YBCO interface. Red arrows indicate local CuO double chains ($Y_{248}$ intergrowths) disrupting the regular twin structure. (d) Simultaneous annular dark field image (ADF) of an Electron Energy Loss Spectroscopy (EELS) map interface. Right: Color mix of the Ba $M_{45}$ (red), Mn $L_{23}$ (green) and La $M_{45}$ (blue) edge signals. **Scale bars:** (c) upper: 100 nm; lower: 10 nm; (d): 1 nm.

signal at the APB plane (red curve) with that of the immediately preceding, defect-free lattice plane (black curve). The APB plane exhibits two distinct maxima of enhanced Z-contrast, corresponding to a double layer of heavy cations (BaO or LaSrO) inserted at the boundary.

Enhanced Z-contrast at this location confirms the presence of additional $Ba^{2+}$/$La^{3+}$ columns relative to the surrounding YBCO matrix. Such anomalous stacking originates from the $MnO_2$-terminated LSMO surface interfacing with the BaO termination of YBCO [48,49]. This termination and the overall structural properties of our samples are similar to those observed in earlier work [40,45,65,66].

The study of microstructural properties of the YBCO(103)/LSMO(110) bilayer heterostructure grown on [100]-oriented NGO included spectroscopic measurements to fully characterize the interface for this crystalline orientation, on which we found no early work in the literature. Figure 3 (c) (upper panel) demonstrates high-quality epitaxial growth, with continuous film coverage and well-defined interfaces at the manganite-superconductor boundaries. The atomically resolved lower panel of Figure 3 (c) (bottom) confirms the sharpness of the LSMO/YBCO interface, with minimal intermixing and coherent stacking [45,67,68]. The most distinctive feature observed in the YBCO layer is the emergence of a characteristic herringbone pattern of {103} twin domains, arising from substrate-induced epitaxial strain during the tetragonal-to-orthorhombic structural phase transition. Unlike YBCO films on cubic substrates (e.g., $SrTiO_3$ or $LaAlO_3$), where twins form parallel lamellae with a fixed ~90° angle between adjacent domains, the orthorhombic symmetry of the [100]-oriented NGO substrate imposes a subtle directional constraint that produces the distinctive crossed herringbone geometry with alternating in-plane orientations and twin spacing ranging from 5 to 15 nm [69,70]. Within this coherent twin network, several regions marked by red arrows in Figure 3 (c) reveal local deviations from the perfect herringbone lattice, originating from planar defects such as copper oxide double chains ($Y_2Ba_4Cu_8O_{16}$, or $Y_{248}$ intergrowths) and, less frequently, triple-chain variants ($Y_2Ba_2Cu_5O_8$, or $Y_{125}$). These intergrowths introduce localized lattice distortions and partial dislocations that partially break the vertical coherence of the twin boundaries, reducing their through-thickness continuity to just a few nanometers. The bending and modulation of lattice planes underscore the intricate interplay between substrate-induced anisotropic strain from the NGO lattice mismatch, interface strain from the LSMO/YBCO boundary, and localized nanostrain fields generated by non-coherent nano-scale defects distributed throughout the YBCO matrix [68,70–72]. Figure 3 (d) displays the interfacial atomic stacking. The growth of the manganite along this direction results in alternating Mn and La/Sr planes at the interface with YBCO, thereby modifying the local chemical environment on the cuprate side. Figure 3 (d) shows the ADF image acquired simultaneously with a STEM Electron Energy Loss Spectroscopy (EELS) chemical map. In the right panel, the color map displays the Ba $M_{4,5}$, La $M_{4,5}$ and Mn $L_{2,3}$ signals across the layer, highlighting the abruptness of the interface and the correlation between the cation sublattices in LSMO and YBCO. In summary, and despite the presence of defects in the bulk of the YBCO layer, such as double or triple-chains that may be relevant for properties such as vortex pinning [70,72], the structural quality of the YBCO/LSMO interface –which is the most relevant property for the physics at stage in the current study– shows a high degree of structural order.

## III. MAGNETIC, SUPERCONDUCTING, AND SPIN PUMPING PROPERTIES.

Typical magnetization loops $M(H)$ for c-axis and (103) samples are shown in Fig. 4. These measurements were done using a SQUID magnetometer, with the field applied in-plane and at T = 100 K, that is, above the superconducting transition. Both samples have similar saturation magnetization ~4.2 $10^5$ A/m. The (103) sample shows a higher saturation field than the c-axis one. This is expected since, for (110) LSMO films, the magnetic hard axis is along the [-110] direction [73,74], which is parallel to the magnetic field (see Fig. 3 for crystalline orientation). For c-axis films, the easy axes lie within the (001) plane [73,75], particularly along [110] and [-110] directions, which are coplanar with the applied field direction and only 45 degrees off it.

Spin pumping was characterized via conventional broadband ferromagnetic resonance (FMR) experiments carried out in a variable temperature PPMS cryostat of Quantum Design equipped with aa 7T electromagnet to apply the dc field $H$. The samples were mounted upside down (YBCO down) on a Coplanar Waveguide (CPW) [see Fig. 1(a)] that allows applying a microwave field $h_{rf}$ of tunable amplitude and frequency $f$ (in the 1 GHz – 40 GHz range). During the experiments, we measured the derivative of the microwave power transmitted across the CPW $dP/dH$ as a function of $H$, using a rf diode and lock-in detection [76]. Fig. 5(a) shows a series of measurements for different frequencies at constant temperature, while Fig. 5(b) shows a measurement for a single frequency, as an example. The microwave absorption line shape corresponds to a Lorentzian, whose derivative is characterized by the peak-to-peak linewidth $\mu_0\Delta H_{pp}$ [marked with the double-headed arrow in Fig. 5(d)] centered around the resonance field $\mu_0 H_r$. The experimental data is fitted with a linear combination of symmetric and antisymmetric functions of the form,

$$\frac{dP}{dH}(H)=A\cdot\frac{2\frac{\Delta H}{2}(H\text{-}H_0)}{\left[(H\text{-}H_0)^2+\left(\frac{\Delta H}{2}\right)^2\right]^2}+B\cdot\frac{\left(\frac{\Delta H}{2}\right)^2\text{-}(H\text{-}H_0)^2}{\left[(H\text{-}H_0)^2+\left(\frac{\Delta H}{2}\right)^2\right]^2} \quad (4)$$

which are calculated from the derivative of the Lorentzian expression [77,78]. Here, $\mu_0 H_r$ is the resonance field and, for the case of a purely asymmetric lineshape (B = 0), the linewidth $\Delta H$ and $\Delta H_{pp}$ are related as $\Delta H=\sqrt{3}\ H_{pp}$.The behavior observed in Fig. 5 is qualitatively similar for all the studied samples. A single resonance is observed for each frequency, and our data show a dominant contribution of the asymmetric part. However, the symmetric contribution is not negligible and we included it for the fitting to get more accurate values of $\mu_0 \Delta H_{pp}$. To calculate the magnetic damping, we measured $\mu_0 \Delta H_{pp}$ for different frequencies, which in the linear regime is proportional to the frequency [21,78] :

$$\mu_0 \Delta H_{pp}=\frac{2\alpha f}{\sqrt{3}\gamma}+\mu_0 \Delta H_0 \quad (5)$$

where $\mu_0 \Delta H_0$ is the frequency-independent inhomogeneous broadening.

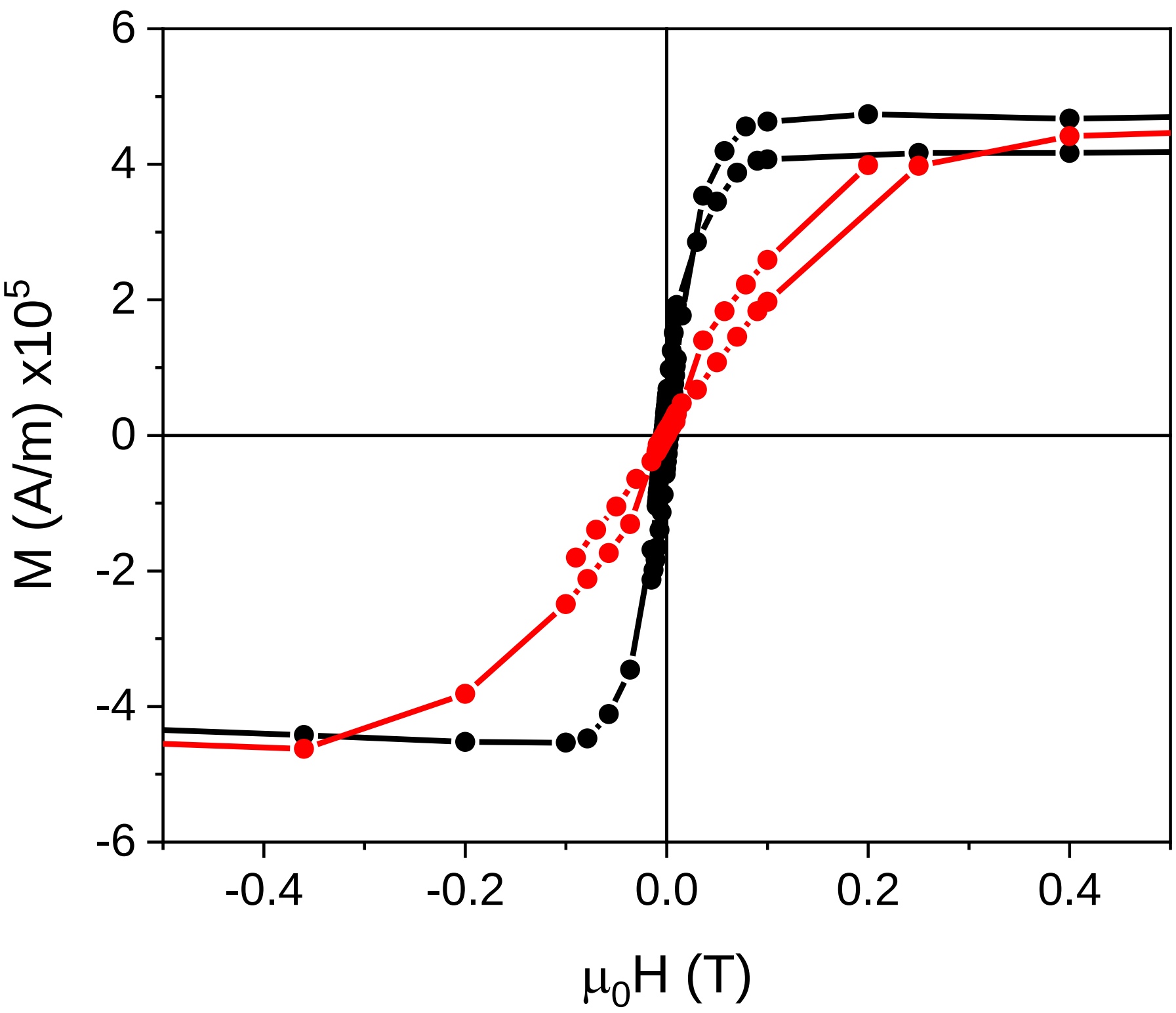

**Fig.4 :** Magnetization vs field for a c-axis bilayer (black) and a (103) bilayer (red). The field was applied in-plane for both cases, similar to the FMR experiments. In this configuration, for the (103) sample the field is applied along the [-110], while for c-axis films it is applied along [010].

The behavior observed in Fig. 5 is the same for all $T$, and in particular above and below the superconducting transition of S samples. From the quantitative standpoint, fitting of the measurement for each frequency using Eq. 4, as shown in Fig. 5(b), allows extracting the frequency-dependent peak-to-peak linewidth $\mu_0 \Delta H_{pp}$ and the resonant field $\mu_0 H_r$. Their relation with the $f$ is displayed in Figs. 5 (c) and 5 (d). We find a canonical behavior, as the resonant field follows the Kittel formula (the red line in Fig. 5 (c) is a data fit to Eq. 2). This allows determination of $\mu_0 H_k$ and $M_{eff}$. On the other hand, the linewidth vs. frequency is linear, and a fit using Eq. 5 [see Fig. 5(d)] allows determination of the damping coefficient $\alpha$ and the inhomogeneous broadening $\mu_0 \Delta H_0$. Such analysis is performed for datasets as those in Fig. 5 (a) taken at different temperatures, to obtain $\alpha$(T), $\mu_0 \Delta H_0$(T), $\mu_0 H_k$ $(T)$ and $M_{eff}$(T). An example of this analysis, accompanied by electrical transport characterization of superconductivity, is shown in Fig. 6, which contains data for a c-axis S sample and its nS counterpart.

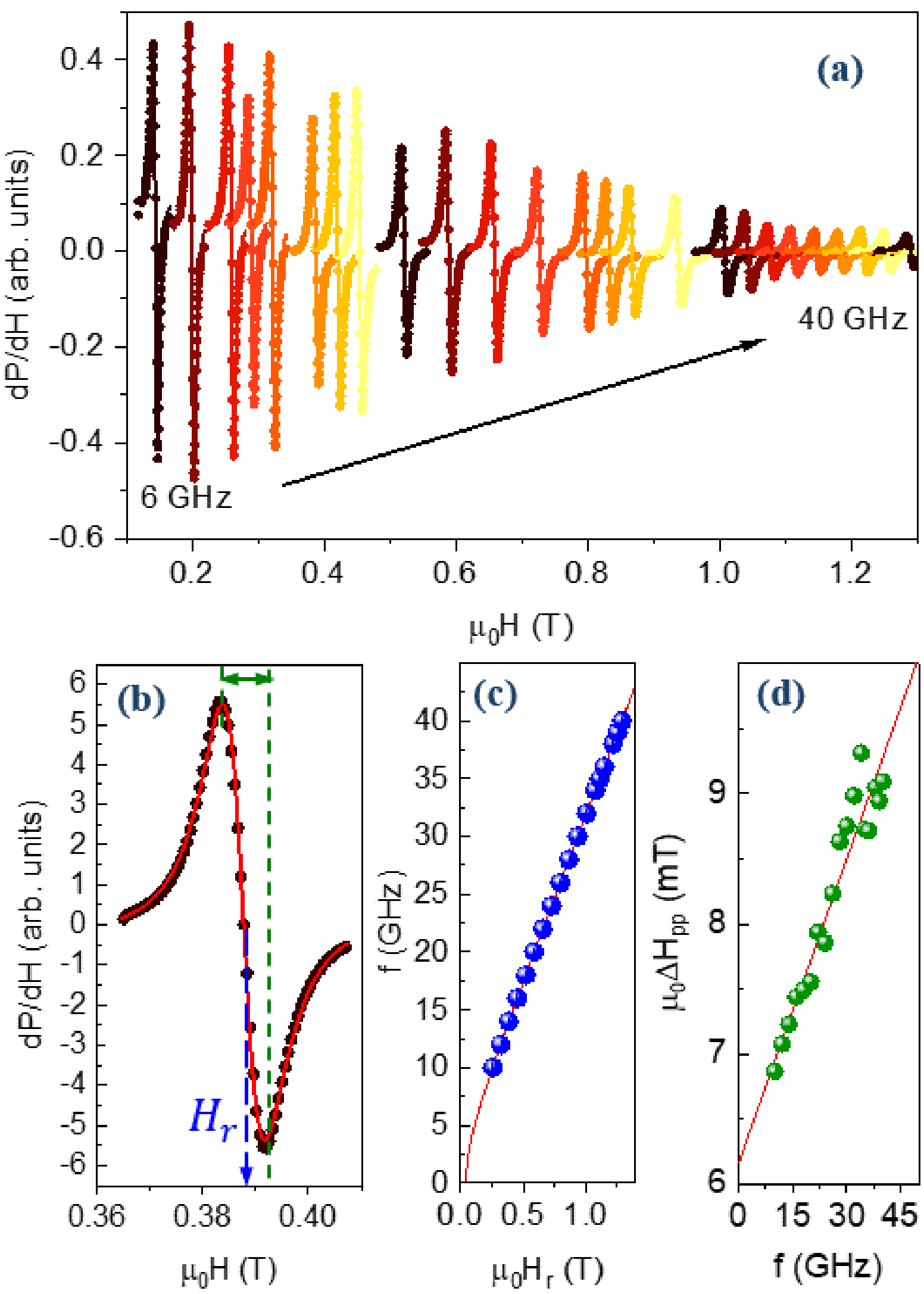

**Fig.5:** (a) Typical FMR spectra obtained for different frequencies at a fixed temperature. The signal size decreases for high frequencies mainly due to the impedance mismatch between the CPW and the sample. (b) Typical FMR spectrum measured at a fixed frequency and temperature, where the peak-to-peak linewidth $\mu_0 \Delta H_{pp}$ (green) and the resonance field $\mu_0 H_r$ (blue) are indicated. (c) Frequency vs $\mu_0 H_r$ for a typical temperature, where the fit is based on the Kittel law Eq. (2). (d) $\mu_0 \Delta H_{pp}$ vs frequency (green dots) and fit using Eq. 5 (red line).

Fig. 6 (a) shows four-probe resistance versus temperature $R(T)$ measurements carried out in the presence of a microwave field and in the absence of a DC magnetic field. While the nS sample shows metallic behavior down to the lowest temperature, with no hint of superconductivity, for the S samples (red line) a resistive transition from the normal into the superconducting state is observed. The superconducting critical temperature $T_c$, defined by the onset of the resistive transition (as indicated in the figure), is of the order of 74 K. In the studied S samples, the $T_c$ was always in the range 74 K-80 K, suggesting that YBCO is slightly underdoped.

Fig. 6 (b) shows the linewidth $\mu_0 \Delta H_{pp}$(T) for different $f$ obtained for the S sample (see legend). For $f$ up to 26 GHz, the linewidth steadily broadens upon cooling down, without any noticeable change through the superconducting transition (marked by a vertical dashed line). However, for higher frequencies, the trend of the linewidth broadening is enhanced below Tc. This departure from the higher-temperature trend is stronger the higher the frequency. Such behavior suggests an additional frequency-dependent contribution to the magnetization relaxation appearing at the onset of the superconducting transition. This is more clearly demonstrated by the analysis below.

Fig. 6 (c) compares $\alpha(T)$ for the S and the reference nS sample. For the reference sample (black symbols), the damping $\alpha(T)$ shows a weak, monotonic temperature dependence across the entire temperature range. The superconducting sample (red symbols) shows a similar trend in the normal state. However, upon cooling below $T_C$ , $\alpha(T)$ grows drastically: at low-temperatures, $\alpha$ reaches as much as twice its normal-state value. In contrast to that, the temperature dependence of $\mu_0 \Delta H_0(T)$, $M_{eff}(T)$ and $\mu_0 H_k(T)$ [see Figs. 6 (d)-(f)] are very similar for the S and nS samples, and none of those three parameters show a drastic change across the superconducting transition. This implies, on the one hand, that the intrinsic magnetic properties of the LSMO layer are similar in the S and nS samples, and on the other hand, that $\mu_0 H_0(T)$, $\mu_0 H_k(T)$ and $M_{eff}(T)$ are unaffected by the superconducting transition. Thus, the abrupt change of the damping coefficient $\alpha$ is the only dramatic effect produced by the onset of superconductivity. The conclusion of all the above is that superconductivity radically changes the spin-sinking behavior of YBCO. For the dataset in Fig. 6, the fact that $\alpha(T)$ increases upon

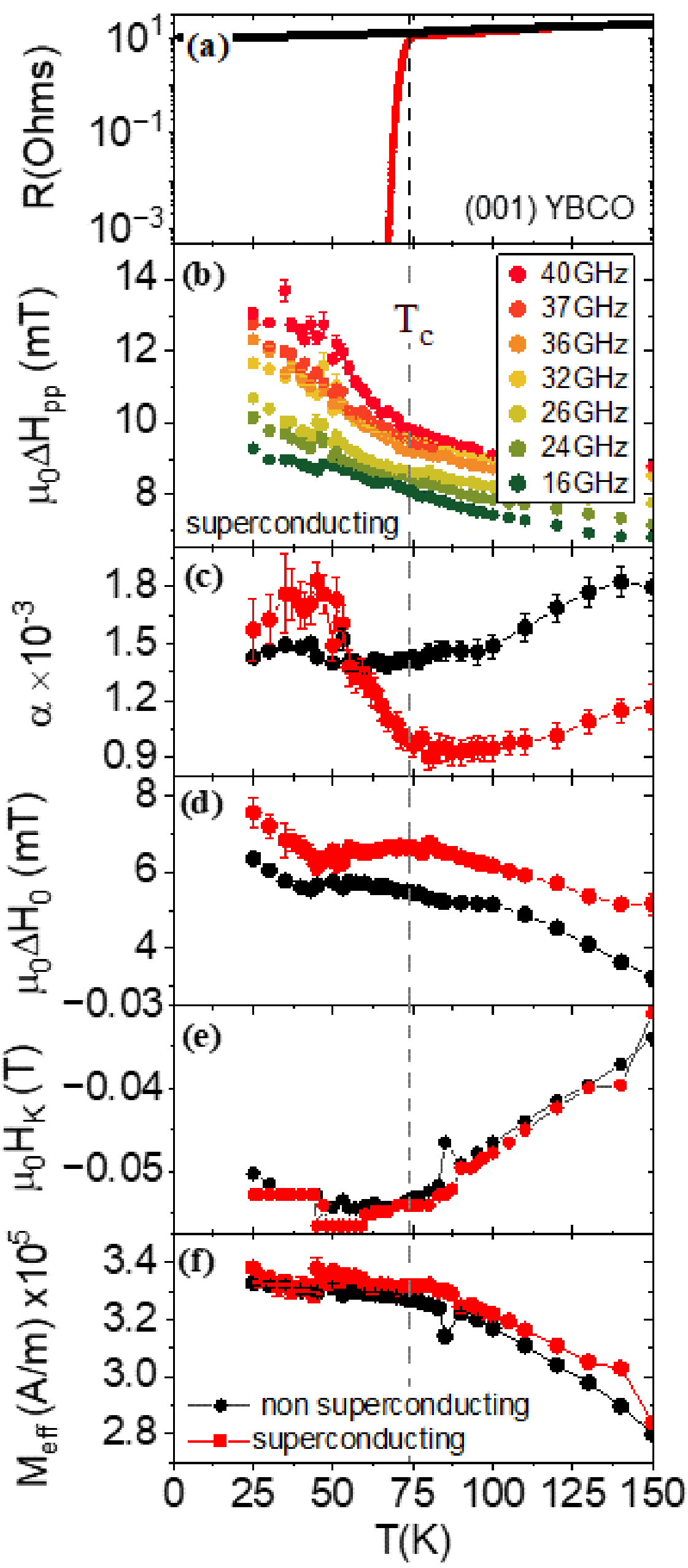

**Fig. 6:** Results obtained as a function of temperature T for a c-axis superconducting bilayer (S) (in red) and its non-superconducting reference (nS) (in black). (a) R vs T, (b) $\mu_0 \Delta H_{pp}$ vs T of the S sample for different frequencies, (c) damping $\alpha$ vs T, (d) inhomogeneous broadening $\mu_0 H_0$ vs T, (e) anisotropy field $\mu_0 H_k$ vs T and effective magnetization $M_{eff}$ vs T

cooling below $T_C$ indicates that spin-pumping (and thus the spin current) into YBCO increases as the material becomes superconducting. To investigate how this unusual behavior depends on the crystalline orientation of the interface, and on the LSMO layer thickness, in what follows we compare $\alpha(T)$ for c-axis sample in Fig. 6 with $\alpha(T)$ for a sample having similar LSMO thickness but grown along the (103) direction, as well as with another two c-axis samples having different LSMO thickness.

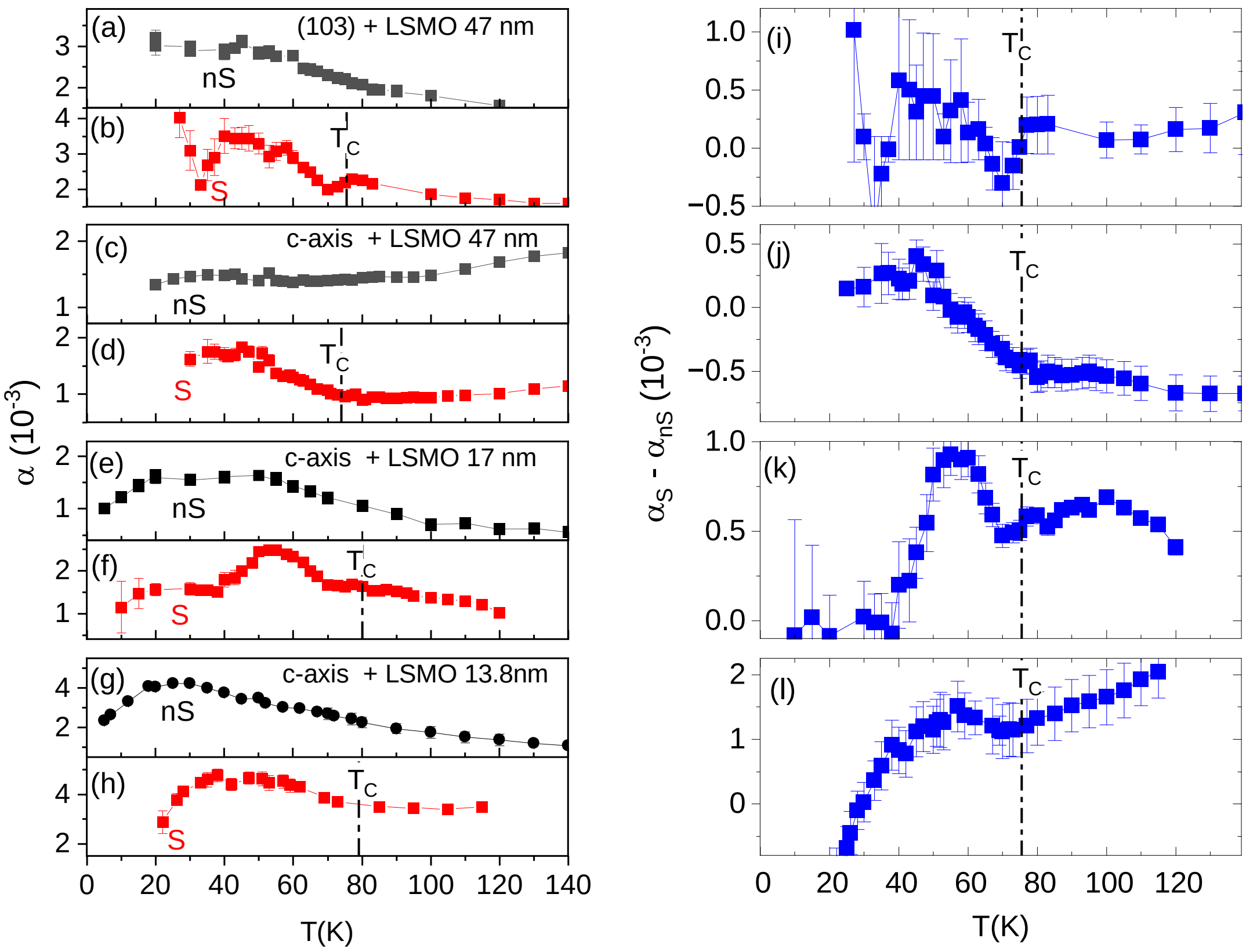

**Fig. 7 :** (a-h) vs T for all the reference nS (in black) and superconducting S (in red) samples. In (a-d) we show the results for the samples with 47 nm thick LSMO and YBCO oriented along the (a-b) (103) direction and (c-d) c-axis direction. (e-f) and (g-h) show vs T for other c-axis samples with LSMO thicknesses of (e-f) 17 nm and (g-h) 13.8 nm. In (i-l) we computed the difference between the damping of the superconducting and non-superconducting samples $\alpha_S$-$\alpha_{nS}$ vs T. The critical temperature Tc is indicated with a vertical dashed line.

Fig. 7 (a)-(h) show $\alpha(T)$ both for the S samples (red data points) and their nS counterparts (black data points). We can see that the nS samples' behavior varies from one to another: while $\alpha$ lies in the $10^{-3}$ range for all of them, their temperature dependence differs. This aspect will be discussed further below, based on earlier studies of $\alpha(T)$ in LSMO films [79]. For now, let us just use $\alpha(T)$ of the nS samples as a reference to identify the superconducting effects. These stand out clearly, despite the relatively strong temperature dependence of the damping coefficient for some of the nS samples. One can see, upon direct comparison of Figs. 7 (a)-(c)- (e)-(g) with Figs. 7 (b)-(d)-(f)-(h), that $\alpha(T)$ for each S sample follows the trend of its nS counterpart for temperatures down to ~ $T_C$. Below this temperature, however, a departure from the nS trend is observed in all cases. This can be more clearly appreciated in Figs. 7 (i) to 7(l), which display the difference between the damping coefficient of each S sample and its nS reference $\Delta\alpha(T)$=$\alpha_S(T)$-$\alpha_{nS}(T)$. The (103) sample [Fig. 7 (i)] shows a clear drop below $T_C$, indicating a relative decrease of spin-pumping at the onset of superconductivity, followed by an increase above the background level upon further temperature decrease. The behavior at the lowest temperatures is unclear as the error bars become comparable to $\Delta\alpha$ variations. For c-axis films, Fig. 7 (j)-(k)-(l), strikingly, we observe no significant drop right below $T_C$, but on the contrary, a marked increase of the damping that indicates enhanced spin-pumping. Upon further cooling, a local maximum is reached, followed by a gradual decay in samples where low temperature measurements are available [Figs. 7 (k) and (l)].

In summary, the experiments unveil a distinct temperature evolution of $\alpha(T)$ at the onset of superconductivity depending on the crystalline orientation of the heterostructures. For the (103) ones, the magnetic damping drastically drops just below $T_C$, which is followed by an increase, similarly as observed earlier for YBCO/Py heterostructures with rough interfaces. However, c-axis heterostructures show an unusual behavior: $\alpha$ increases below $T_C$, reaching a maximum at ~0.65–0.7$T_C$, before decreasing again upon further temperature decrease. This contrasts with the expectations considering experiments on s-wave [25,80] and c-axis d-wave superconductors with flat interfaces [33], based on which we would have anticipated a drastic drop in the damping just below $T_C$ due to the opening of the superconducting gap.

## IV. DISCUSSION

Let us start by discussing the behavior of the nS samples. For these, we expect that the temperature dependence of the Gilbert damping is dominated by the intrinsic behavior of the underlying LSMO film. This is due to the poorly metallic nature of the 5-nm-thick YBCO overlayer (with the top ~2 nm being insulating as a result of deoxygenation [81]), whose electronic and magnetic properties (including spin diffusion length) are expectedly weakly temperature dependent. Based on this, spin-pumping into the YBCO overlayer is anticipated to contribute only weakly to the temperature dependence of the FMR linewidth and, particularly, to the temperature dependence of $\alpha$. Indeed, $\alpha(T)$ of our c-axis LSMO films [Fig. 7(c), (e), and (g)] is fully consistent with that observed in recent studies of $\alpha(T)$ in plain c-axis LSMO films [79,82]. For the thicker LSMO sample [47 nm, Fig. 7(c)], $\alpha(T)$ decays with decreasing temperature until a weak enhancement is observed below ~40 K. This behavior indicates a dominance of the so-called "resistively-like" term in the Gilbert damping, which is proportional to the temperature-dependent resistivity and is related to inter-band scattering processes that involve the excitation of electron-hole pairs across different bands [79]. The dominance of inter-band scattering is intrinsically expected for half-metals, as opposed to the intra-band scattering involving spin-↑ to spin-↓ transitions that are strongly hindered in ideal conditions [79]. In the literature, an inter-band scattering dominance is observed in c-axis films thicker than ~30 nm [82] with *H* applied in-plane; therefore, our c-axis 47 nm film behaves as expected. On the other hand, experiments have shown [79] that, when the thickness of c-axis LSMO films is reduced below ~ 20 nm, such "bulklike" behavior is lost: $\alpha$ is largely enhanced and steadily increases upon decreasing $T$, with an eventual drop below ~ 30 K. This characteristic behavior is more pronounced the thinner the LSMO. This is as observed in our thin c-axis films, see Figs. 7 (e) and (g). This phenomenology has been explained considering the presence of a phase-separated "dead" layer that is magnetically active and presents a shorter spin relaxation time than the bulk of the manganite film, therefore acting as a spin sink and enhancing $\alpha$, with a characteristic temperature dependence that is related to the freezing of spin fluctuations [79]. Finally, although we could not find earlier literature on FMR of (110) LSMO film, the high $\alpha$ as compared with that of the c-axis film of similar thickness and its steady increase with decreasing temperature [Fig. 7 (a)] suggests the dominance of intra-band scattering. This is proportional to the conductivity [79] and dominates in the presence of canting between $t_{2g}$ magnetic moments, as this yields a non-zero density of minority spin-↓ states at the Fermi level. Canting of the Mn moments is indeed expected for our (110) LSMO film because $H$ is applied parallel to its magnetic hard axis [-110] (see Fig. 4 and [73,74]).

Let us now discuss the behavior of S samples. Despite the very characteristic and marked temperature dependence of the nS samples depending on their thickness and crystalline orientation, we observe that, for all their S counterparts, the onset of superconducting transition dramatically affects $\alpha(T)$, indicating a change in spin-pumping.

For the (103) sample, $\Delta\alpha(T)$ shows a marked drop just below $T_C$, followed by an increase upon further cooling, see Fig. 7 (i). This behavior is reminiscent of c-axis YBCO/Py bilayers with rough interfaces [33], in particular YBCO films with surface crystallites that expose the YBCO's *ab* plane and the Py layer. In that case, $\alpha(T)$ was explained considering two channels for spin pumping in the superconducting state. The first channel is conventional zero-energy QPs created by thermal excitation. The second channel, characteristic of d-wave superconductivity, results from zero-energy surface-bound states (nodal QPs) that appear when the crystalline orientation is such that the *ab* plane is not parallel to the YBCO surface [83–85]. The former channel explains the drop in $\alpha(T)$ below Tc, similarly as for s-wave superconductors [25,80], since the population of thermal QP diminishes as the temperature is decreased below $T_C$. In contrast, the density of the nodal QPs (specific to d-wave pairing) monotonically grows upon cooling below $T_C$ [33,83–85]. That is, spin-pumping via nodal QPs is gradually enhanced as temperature is decreased. Thus, analogously to c-axis films with rough surfaces [33], the coexistence of those two channels can explain the behavior of $\alpha(T)$ for the (103) sample, in which the *ab* plane is not parallel to the YBCO surface. Notice that neither of those two channels requires the occurrence of the superconducting proximity effect, that is, the leakage of superconducting correlations into the ferromagnet.

Let us now turn to the c-axis YBCO/LSMO heterostructures, which show the unexpected behavior that is the central result of this paper. Given the very smooth interfaces [Fig. 3 (a)-(b)], one could have anticipated a behavior similar to that of c-axis YBCO/Py heterostructures with smooth interfaces [33], characterized by a drastic damping decrease below $T_C$ followed by a saturation below the normal-state level –indeed as observed for standard s-wave systems [25,80]. The trends in Fig. 7 (j)-(k)-(l) strongly differ from that. Instead, they are reminiscent of the theoretical predictions in Ref. [29] (see Figs. 3 (a)-(d) in that paper), which considers an s-wave superconductor/ferromagnetic-insulator interface. We observe in Figs. 7 (j)-(k)-(l) that the damping increases below $T_C$, reaching a maximum at around $0.65T_C$ – $0.7T_C$, and subsequently drops upon further cooling. The theoretical model [29] considers that the superconducting order pairing is strongly depressed at the interface relative to the bulk of the superconductor [29]. This inhomogeneity of the order parameter leads to interfacial Andreev bound states, yielding a local enhancement of the density of states available for spin absorption. This explains the enhancement of spin pumping upon cooling below $T_C$, when those bound states form. The temperature dependence of $\alpha(T)$, and particularly the fact that it reaches a maximum and subsequently decreases, is linked to the temperature-dependent occupation of those states, and strongly depends on the thickness $d_N$ of the interfacial slab where those states reside [29]: while the peak appears close to $T_C$ if the slab thickness $d$ is negligible as compared to the coherence length $\xi_0$, when these two length scales are comparable, the spin-pumping maximum can be shifted to temperatures as low as ~ $0.1T_C$. The $\alpha(T)$ maxima in Fig. 7 (j) – (k) – (l), observed around $0.65T_C$-$0.7T_C$, imply $d/\xi_0$ ~ 0.2 according to the model calculations [29].

We discuss in what follows why the model in [29] captures the behavior of the c-axis YBCO/LSMO heterostructures. First, although this model considers an s-wave superconductor, it applies to our experiments because d-wave pairing effects are not relevant for spin pumping in c-axis heterostructures with flat interfaces [33]. This is because, when probed along the c-axis, YBCO's zero-energy density of states shows a temperature dependence similar to s-wave systems [33,34,85]. Second, the fact that our experiments do not involve a ferromagnetic insulator should not substantially affect the qualitative predictions of the model −which can be generalized to the case of metallic ferromagnets by introducing the finite spin-dependent tunneling probability across the interface [29]. The cornerstone of the model is indeed the presence of interfacial Andreev bound states that locally enhance the zero-energy density-of-states, boosting spin sinking. Those states are indeed naturally expected in c-axis LSMO/YBCO, considering the superconducting proximity effect demonstrated in these heterostructures [45,53,54], which is mediated by the interfacial generation of equal-spin Andreev pairs [41,44]. Furthermore, earlier models of atomic-scale superconductor/metallic ferromagnet superlattices [87] and superconductor/normal metal multilayers [88,89] generally predict a strong increase in the density-of-states within the proximitized layers due to zero-energy Andreev states [87–89]. These atomic-scale models are consistently pertinent considering that the spin-pumping maxima at $0.65T_C$-$0.7T_C$ suggest, according to the model [29], that the Andreev bound states reside within a short distance from the interface $d\sim0.2\xi_0 <$ 1 nm. All of that supports the interpretation of our experimental results within the framework of [29]. Notice, in contrast, that the analysis of "bulk" superconducting/ferromagnetic layers (i.e., considering length scales further away from the interface) shows the conventional, significant decrease in the electronic zero-energy in the superconductor near the Fermi level [90]. This explains the markedly different temperature behavior of Py/Nb bilayers, where a damping decrease was observed across the superconducting transition [18], as well as in c-axis YBCO/Py [33] and YBCO(103)/LSMO(110) interfaces, where the superconducting proximity effect has not been observed.

## V. CONCLUSIONS

We have experimentally investigated spin-pumping in half-metallic ferromagnet (LSMO)/d-wave superconductor (YBCO) heterostructures. A thorough structural characterization allows us to ascertain the structural quality of heterostructures grown on different substrates to obtain different crystalline orientations. The ferromagnetic resonance experiments reveal an anomalous behavior of $\alpha(T)$ depending on that. For the (103) YBCO/LSMO heterostructures, $\alpha(T)$ exhibits a drastic drop below $T_C$, followed by an increase to nearly or above normal-state levels. This suggests spin transport via two channels: on the one hand, thermal quasiparticles, which dominate across the superconducting transition and explain the damping drop below $T_C$, and on the other, nodal quasiparticles, which are accessible because the YBCO's *ab* plane is exposed to the interface and play a dominant role in the damping recovery at lower temperatures. At variance, for c-axis YBCO/LSMO, $\alpha(T)$ initially increases below $T_C$, peaking at ~0.65–0.7$T_C$ before decaying for lower temperatures. Such behavior is in stark contrast to experiments with conventional s-wave systems and c-axis YBCO/metallic ferromagnet heterostructures with smooth interfaces, where $\alpha(T)$ typically drops right below $T_C$. The behavior is as theoretically predicted in the presence of interfacial Andreev bound states [29], which are expected in the c-axis LSMO/YBCO interface because the superconducting proximity effect is strong. In summary, our study illustrates how spin-pumping into superconductors can be enhanced by different types of bound quasiparticles: on the one side, nodal quasiparticles, specific to d-wave superconductivity, and on the other, quasiparticles emerging due to spatial inhomogeneities of the order parameter, which are not specific to d-wave superconductivity and become dominant in our c-axis heterostructures due to the superconducting proximity effect.

## ACKNOWLEDGMENTS

We acknowledge funding from French ANR through Grants No. ANR-22-CE30-00020-01 “SUPERFAST,” No. PEPR SPIN ANR-22-EXSP-0007 “SPINMAT,” and No. ANR-24-EXSP-0012 “SUPERSPIN”; the European EIC pathfinder Grant No. 101130224 “JOSEPHINE”; Cost Action CA21144 “SUPERQUMAP”; Agencia Estatal de Investigación grants PID2023-148884OB-I00 and TED 2021130196B-C21; “(MAD2D-CM)-UCM” project funded by Comunidad de Madrid, by the Recovery, Transformation and Resilience Plan, and by Next Generation EU from the European Union. We thank the ICTS-ELECMI CNME node for access to its facilities. We thank A. Anane (Laboratoire Albert Fert) for discussions and insights.

[1] A. C. Basaran, J. E. Villegas, J. S. Jiang, A. Hoffmann, and I. K. Schuller, Mesoscopic magnetism and superconductivity, MRS Bull. **40**, 925 (2015).
[2] I. F. Lyuksyutov and V. L. Pokrovsky, Ferromagnet–superconductor hybrids, Adv. Phys. **54**, 67 (2005).
[3] A. A. Golubov, S. V. Bakurskiy, M. Yu. Kupriyanov, T. Karabassov, A. S. Vasenko, and A. S. Sidorenko, The physics of superconductor-ferromagnet hybrid structures, (2025).
[4] O. Dobrovolskiy et al., Roadmap on nanoscale superconductivity for quantum technologies, Supercond. Sci. Technol. **39**, 023502 (2026).
[5] A. I. Buzdin, Proximity effects in superconductor-ferromagnet heterostructures, Rev. Mod. Phys. **77**, 935 (2005).
[6] F. S. Bergeret, A. F. Volkov, and K. B. Efetov, Odd triplet superconductivity and related phenomena in superconductor- ferromagnet structures, Rev. Mod. Phys. **77**, 1321 (2005).
[7] X. Palermo et al., Tailored Flux Pinning in Superconductor-Ferromagnet Multilayers with Engineered Magnetic Domain Morphology from Stripes to Skyrmions, Phys. Rev. Appl. **13**, 1 (2020).
[8] D. Sanchez-Manzano et al., Size-Dependence and High Temperature Stability of Radial Vortex Magnetic Textures Imprinted by Superconductor Stray Fields, ACS Appl. Mater. Interfaces **16**, 19681 (2024).
[9] S. Takahashi, H. Imamura, and S. Maekawa, Spin Imbalance and Magnetoresistance in Ferromagnet/Superconductor/Ferromagnet Double Tunnel Junctions, Phys. Rev. Lett. **82**, 3911 (1999).
[10] N. Poli, J. Morten, M. Urech, A. Brataas, D. Haviland, and V. Korenivski, Spin Injection and Relaxation in a Mesoscopic Superconductor, Phys. Rev. Lett. **100**, 136601 (2008).
[11] C. H. L. Quay, D. Chevallier, C. Bena, and M. Aprili, Spin imbalance and spin-charge separation in a mesoscopic superconductor, Nat. Phys. **9**, 84 (2013).
[12] H. Yang, S.-H. Yang, S. Takahashi, S. Maekawa, and S. S. P. Parkin, Extremely long quasiparticle spin lifetimes in superconducting aluminium using MgO tunnel spin injectors., Nat. Mater. **9**, 586 (2010).
[13] O. V. Dobrovolskiy, R. Sachser, T. Brächer, T. Böttcher, V. V. Kruglyak, R. V. Vovk, V. A. Shklovskij, M. Huth, B. Hillebrands, and A. V. Chumak, Magnon–fluxon interaction in a ferromagnet/superconductor heterostructure, Nature Physics 2019 15:5 **15**, 477 (2019).
[14] O. V. Dobrovolskiy, Q. Wang, D. Y. Vodolazov, R. Sachser, M. Huth, S. Knauer, and A. I. Buzdin, Moving Abrikosov vortex lattices generate sub-40-nm magnons, Nature Nanotechnology 2025 20:12 **20**, 1764 (2025).
[15] S. V. Mironov, A. S. Mel'nikov, and A. I. Buzdin, Photogalvanic phenomena in superconductors supporting intrinsic diode effect, Phys. Rev. B **109**, L220503 (2024).
[16] Y. Yao, Q. Song, Y. Takamura, J. P. Cascales, W. Yuan, Y. Ma, Y. Yun, X. C. Xie, J. S. Moodera, and W. Han, Probe of spin dynamics in superconducting NbN thin films via spin pumping, Phys. Rev. B **97**, 224414 (2018).
[17] K.-R. Jeon, C. Ciccarelli, A. J. Ferguson, H. Kurebayashi, L. F. Cohen, X. Montiel, M. Eschrig, J. W. A. Robinson, and M. G. Blamire, Enhanced spin pumping into superconductors provides evidence for superconducting pure spin currents, Nat. Mater. **17**, 499 (2018).
[18] C. Bell, S. Milikisyants, M. Huber, and J. Aarts, Spin Dynamics in a Superconductor-Ferromagnet Proximity System, Phys. Rev. Lett. **100**, 47002 (2008).

[19] K.-R. Jeon, J.-C. Jeon, X. Zhou, A. Migliorini, J. Yoon, and S. S. P. Parkin, Giant Transition-State Quasiparticle Spin-Hall Effect in an Exchange-Spin-Split Superconductor Detected by Nonlocal Magnon Spin Transport, ACS Nano **14**, 15874 (2020).
[20] J. Linder and J. W. A. Robinson, Superconducting spintronics, Nat. Phys. **11**, 307 (2015).
[21] T. L. Gilbert, A phenomenological theory of damping in ferromagnetic materials, IEEE Trans. Magn. **40**, 3443 (2004).
[22] C. Kittel, On the Theory of Ferromagnetic Resonance Absorption, Physical Review **73**, 155 (1948).
[23] S. Takahashi, *Physical Principles of Spin Pumping BT - Handbook of Spintronics*, in edited by Y. Xu, D. D. Awschalom, and J. Nitta (Springer Netherlands, Dordrecht, 2016), pp. 1445–1480.
[24] M. Inoue, M. Ichioka, and H. Adachi, Spin pumping into superconductors: A new probe of spin dynamics in a superconducting thin film, Phys. Rev. B **96**, 024414 (2017).
[25] C. Pfaff, S. Petit-Watelot, S. Andrieu, L. Pasquier, J. Ghanbaja, S. Mangin, K. Dumesnil, and T. Hauet, Spin injection at MgB2-superconductor/ferromagnet interface, Appl. Phys. Lett. **125**, (2024).
[26] M. Tinkham, *Introduction to Superconductivity*, 2nd editio (Dover Publication Inc., New York, USA, 2004).
[27] T. Kato, Y. Ohnuma, M. Matsuo, J. Rech, T. Jonckheere, and T. Martin, Microscopic theory of spin transport at the interface between a superconductor and a ferromagnetic insulator, Phys. Rev. B **99**, 1 (2019).
[28] K. R. Jeon, J. C. Jeon, X. Zhou, A. Migliorini, J. Yoon, and S. S. P. Parkin, Giant transition-state quasiparticle spin-Hall effect in an exchange-spin-split superconductor detected by nonlocal magnon spin transport, ACS Nano **14**, 15874 (2020).
[29] M. A. Silaev, Large enhancement of spin pumping due to the surface bound states in normal metal/superconductor structures, Phys. Rev. B **102**, 180502 (2020).
[30] K.-R. Jeon, C. Ciccarelli, A. J. Ferguson, H. Kurebayashi, L. F. Cohen, X. Montiel, M. Eschrig, J. W. A. Robinson, and M. G. Blamire, Enhanced spin pumping into superconductors provides evidence for superconducting pure spin currents, Nat. Mater. **17**, 499 (2018).
[31] K. R. Jeon, C. Ciccarelli, H. Kurebayashi, L. F. Cohen, X. Montiel, M. Eschrig, S. Komori, J. W. A. Robinson, and M. G. Blamire, Exchange-field enhancement of superconducting spin pumping, Phys. Rev. B **99**, 024507 (2019).
[32] X. Montiel and M. Eschrig, Generation of pure superconducting spin current in magnetic heterostructures via nonlocally induced magnetism due to Landau Fermi liquid effects, Phys. Rev. B **98**, 104513 (2018).
[33] S. J. Carreira, D. Sanchez-Manzano, M. W. Yoo, K. Seurre, V. Rouco, A. Sander, J. Santamaría, A. Anane, and J. E. Villegas, Spin pumping in d -wave superconductor-ferromagnet hybrids, Phys. Rev. B **104**, 144428 (2021).
[34] C. Sun and J. Linder, Spin pumping from a ferromagnetic insulator to an unconventional superconductor with interfacial Andreev bound states, Phys. Rev. B **107**, 144504 (2023).
[35] C. Visani et al., Symmetrical interfacial reconstruction and magnetism in La_{0.7}Ca_{0.3}MnO_{3}/YBa_{2}Cu_{3}O_{7}/La_{0.7}Ca_{0.3}MnO_{3} heterostructures, Phys. Rev. B **84**, 060405 (2011).

[36] J. Hoppler et al., Giant superconductivity-induced modulation of the ferromagnetic magnetization in a cuprate–manganite superlattice, Nature Materials 2009 8:4 **8**, 315 (2009).
[37] S. Komori, A. Di Bernardo, A. I. Buzdin, M. G. Blamire, and J. W. A. Robinson, Magnetic Exchange Fields and Domain Wall Superconductivity at an All-Oxide Superconductor-Ferromagnet Insulator Interface, Phys. Rev. Lett. **121**, 077003 (2018).
[38] A. Di Bernardo, S. Komori, G. Livanas, G. Divitini, P. Gentile, M. Cuoco, and J. W. A. Robinson, Nodal superconducting exchange coupling, Nat. Mater. **18**, 1194 (2019).
[39] A. Hoffmann, S. G. E. te Velthuis, Z. Sefrioui, J. Santamaría, M. R. Fitzsimmons, S. Park, and M. Varela, Suppressed magnetization in La0.7Ca0.3MnO3/YBa2Cu3O7−$\delta$ superlattices, Phys. Rev. B **72**, 140407 (2005).
[40] Z. Sefrioui, D. Arias, V. Peña, J. E. Villegas, M. Varela, P. Prieto, C. León, J. L. Martinez, and J. Santamaria, Ferromagnetic/superconducting proximity effect in La0.7Ca0.3MnO3/YBa2Cu3O7−δ superlattices, Phys. Rev. B **67**, 214511 (2003).
[41] C. Visani, Z. Sefrioui, J. Tornos, C. Leon, J. Briatico, M. Bibes, A. Barthélémy, J. Santamaría, and J. E. Villegas, Equal-spin Andreev reflection and long-range coherent transport in high-temperature superconductor/half-metallic ferromagnet junctions, Nat. Phys. **8**, 539 (2012).
[42] Y. Kalcheim, O. Millo, M. Egilmez, J. W. A. Robinson, and M. G. Blamire, Evidence for anisotropic triplet superconductor order parameter in half-metallic ferromagnetic La_{0.7}Ca_{0.3}Mn_{3}O proximity coupled to superconducting Pr_{1.85}Ce_{0.15}CuO_{4}, Phys. Rev. B **85**, 104504 (2012).
[43] J. W. A. Robinson, F. Chiodi, M. Egilmez, G. B. Halász, and M. G. Blamire, Supercurrent enhancement in Bloch domain walls, Scientific Reports 2012 2:1 **2**, 1 (2012).
[44] C. Visani, F. Cuellar, A. Pérez-Muñoz, Z. Sefrioui, C. León, J. Santamaría, and J. E. Villegas, Magnetic field influence on the proximity effect at YB a2 C u3 O7/ L a2/3 C a1/3Mn O3 superconductor/half-metal interfaces, Phys. Rev. B **92**, 014519 (2015).
[45] D. Sanchez-Manzano et al., Extremely long-range, high-temperature Josephson coupling across a half-metallic ferromagnet, Nat. Mater. **21**, 188 (2022).
[46] D. Sanchez-Manzano et al., Unconventional long range triplet proximity effect in planar YBa2Cu3O7/La0.7Sr0.3MnO3/YBa2Cu3O7 Josephson junctions, Supercond. Sci. Technol. **36**, 074002 (2023).
[47] J. Salafranca and S. Okamoto, Unconventional Proximity Effect and Inverse Spin-Switch Behavior in a Model Manganite-Cuprate-Manganite Trilayer System, Phys. Rev. Lett. **105**, 256804 (2010).
[48] C. L. Prajapat, S. Singh, D. Bhattacharya, G. Ravikumar, S. Basu, S. Mattauch, J. G. Zheng, T. Aoki, and A. Paul, Proximity effects across oxide-interfaces of superconductor-insulator-ferromagnet hybrid heterostructure, Scientific Reports 2018 8:1 **8**, 3732 (2018).
[49] B. A. Gray et al., Superconductor to Mott insulator transition in YBa2Cu3O7/LaCaMnO3 heterostructures, Scientific Reports 2016 6:1 **6**, 33184 (2016).
[50] A. Lagarrigue et al., Bipolar spin-valve effect in complex-oxide superconductor/half-metallic ferromagnet junctions, Phys. Rev. Mater. **10**, 015002 (2026).
[51] V. Peña, Z. Sefrioui, D. Arias, C. Leon, J. Santamaria, J. L. Martinez, S. G. E. te Velthuis, and A. Hoffmann, Giant Magnetoresistance in Ferromagnet/Superconductor Superlattices, Phys. Rev. Lett. **94**, 57002 (2005).

[52] R. de Andrés Prada, T. Golod, O. M. Kapran, E. A. Borodianskyi, Ch. Bernhard, and V. M. Krasnov, Memory-functionality superconductor/ferromagnet/superconductor junctions based on the high-$Tc$ cuprate superconductors YBa2⊠Cu3⊠O7−$x$ and the colossal magnetoresistive manganite ferromagnets La2/3⊠$X$1/3⊠MnO3+$\delta$⊠($X$=Ca,Sr), Phys. Rev. B **99**, 214510 (2019).

[53] D. Sanchez-Manzano et al., Unconventional long range triplet proximity effect in planar YBa2Cu3O7/La0.7Sr0.3MnO3/YBa2Cu3O7 Josephson junctions, Supercond. Sci. Technol. **36**, 074002 (2023).

[54] D. Sanchez-Manzano et al., Long-range superconducting proximity effect in YBa2Cu3O7/La0.7Ca0.3MnO3 weak-link arrays, Appl. Phys. Lett. **124**, (2024).

[55] J. Konopka and I. Wolff, Dielectric properties of substrates for deposition of high-T/sub c/ thin films up to 40 GHz, IEEE Trans. Microw. Theory Tech. **40**, 2418 (1992).

[56] Z. Sefrioui, D. Arias, C. Leon, J. Santamaria, and V. Pen, Giant Magnetoresistance in Ferromagnet / Superconductor Superlattices, **057002**, 3 (2005).

[57] R. De Andrés Prada, T. Golod, O. M. Kapran, E. A. Borodianskyi, C. Bernhard, and V. M. Krasnov, Memory-functionality superconductor/ferromagnet/superconductor junctions based on the high-<math xmlns="http://www.w3.org/1998/Math/MathML"><msub><mi>T</mi><mi>c</mi></msub></math> cuprate superconductors <math xmlns="http://www.w3.org/1998/Math/MathML">…, Phys. Rev. B **99**, 214510 (2019).

[58] C. Visani, F. Cuellar, A. Pérez-Muñoz, Z. Sefrioui, C. León, J. Santamaría, and J. E. Villegas, Magnetic field influence on the proximity effect at $\mathrm{YB}{\mathrm{a}}_{2}\mathrm{C}{\mathrm{u}}_{3}{\mathrm{O}}_{7}/\mathrm{L}{\mathrm{a}}_{2/3}\mathrm{C}{\mathrm{a}}_{1/3}\mathrm{Mn}{\mathrm{O}}_{3}$ superconductor/half-metal interfaces, Phys. Rev. B **92**, 14519 (2015).

[59] S. M. Morley, R. P. Campion, K. Horbelt, P. J. King, H. .-U. Habermeier, and B. Leibold, In-plane anisotropic properties of (103)/(013) and 10/spl deg/ YBCO thin films, IEEE Transactions on Applied Superconductivity **7**, 3458 (1997).

[60] J. Johansson, K. Cedergren, T. Bauch, and F. Lombardi, Properties of inductance and magnetic penetration depth in (103)-oriented ${\text{YBa}}_{2}{\text{Cu}}_{3}{\text{O}}_{7\ensuremath{-}\ensuremath{\delta}}$ thin films, Phys. Rev. B **79**, 214513 (2009).

[61] F. Tafuri, F. Miletto Granozio, F. Carillo, A. Di Chiara, K. Verbist, and G. Van Tendeloo, Microstructure and Josephson phenomenology in 45\ifmmode^\circ\else\textdegree\fi{} tilt and twist ${\mathrm{YBa}}_{2}{\mathrm{Cu}}_{3}{\mathrm{O}}_{7\mathrm{\ensuremath{-}}\mathrm{\ensuremath{\delta}}}$ artificial grain boundaries, Phys. Rev. B **59**, 11523 (1999).

[62] R. W. . James, *The Optical Principles of the Diffraction of X-Rays* (Ox Bow Press, 1982).

[63] V. K. Malik et al., Pulsed laser deposition growth of heteroepitaxial YBa<span class="aps-inline-formula"><math xmlns="http://www.w3.org/1998/Math/MathML" display="inline"><msub><mrow></mrow><mn>2</mn></msub></math></span>Cu<span class="aps-inline-formula"><math xmlns="http…, Phys. Rev. B **85**, 054514 (2012).

[64] Z. Wang et al., Designing antiphase boundaries by atomic control of heterointerfaces, Proc. Natl. Acad. Sci. U. S. A. **115**, 9485 (2018).

[65] A. Hoffmann, S. G. E. Te Velthuis, Z. Sefrioui, J. Santamaría, M. R. Fitzsimmons, S. Park, and M. Varela, Suppressed magnetization in <span class="aps-inline-formula"><math

xmlns="http://www.w3.org/1998/Math/MathML" display="inline"><mrow><msub><mi>La</mi><mn>0.7</mn></msub><msub><mi>Ca</mi><mn>0.3</mn></msub><msub><mi>MnO</mi><mn>3</mn></msub><mo>/</mo><msu…, Phys. Rev. B **72**, 140407 (2005).

[66] M. Varela, A. R. Lupini, S. J. Pennycook, Z. Sefrioui, and J. Santamaria, Nanoscale analysis of YBa2Cu3O7-x/La 0.67Ca0.33MnO3 interfaces, Solid. State. Electron. **47**, 2245 (2003).

[67] M. Varela, A. R. Lupini, S. J. Pennycook, Z. Sefrioui, and J. Santamaria, Nanoscale analysis of YBa2Cu3O7-x/La 0.67Ca0.33MnO3 interfaces, Solid. State. Electron. **47**, 2245 (2003).

[68] Y. P. Ivanov, S. Soltan, J. Albrecht, E. Goering, G. Schütz, Z. Zhang, and A. Chuvilin, The Route to Supercurrent Transparent Ferromagnetic Barriers in Superconducting Matrix, ACS Nano **13**, 5655 (2019).

[69] C. J. Jou and J. Washburn, Formation of coherent twins in YBa2Cu3O7–δ superconductors, Journal of Materials Research 1989 4:4 **4**, 795 (2011).

[70] V. Rouco, A. Palau, R. Guzman, J. Gazquez, M. Coll, X. Obradors, and T. Puig, Role of twin boundaries on vortex pinning of CSD YBCO nanocomposites, Supercond. Sci. Technol. **27**, 125009 (2014).

[71] H. Zhang, N. Gauquelin, G. A. Botton, and J. Y. T. Wei, Attenuation of superconductivity in manganite/cuprate heterostructures by epitaxially-induced CuO intergrowths, Appl. Phys. Lett. **103**, (2013).

[72] A. Llordés et al., Nanoscale strain-induced pair suppression as a vortex-pinning mechanism in high-temperature superconductors, Nature Materials 2012 11:4 **11**, 329 (2012).

[73] Y. Suzuki, H. Y. Hwang, S. W. Cheong, and R. B. Van Dover, The role of strain in magnetic anisotropy of manganite thin films, Appl. Phys. Lett. **71**, 140 (1997).

[74] V. Bhosle, J. T. Prater, and J. Narayan, Anisotropic magnetic properties in [110] oriented epitaxial La0.7 Sr0.3 Mn O3 films on (0001) sapphire, J. Appl. Phys. **102**, 13527 (2007).

[75] K. Steenbeck, R. Hiergeist, A. Revcolevschi, and L. Pinsard-Gaudart, Magnetic Anisotropy in La0.7(Sr,Ca)0.3MnO3 Epitaxial Thin Films And Crystals, MRS Online Proceedings Library **562**, 57 (1999).

[76] S. S. Kalarickal, P. Krivosik, M. Wu, C. E. Patton, M. L. Schneider, P. Kabos, T. J. Silva, and J. P. Nibarger, Ferromagnetic resonance linewidth in metallic thin films: Comparison of measurement methods, J. Appl. Phys. **99**, 093909 (2006).

[77] M. Harder, Z. X. Cao, Y. S. Gui, X. L. Fan, and C.-M. Hu, Analysis of the line shape of electrically detected ferromagnetic resonance, Phys. Rev. B **84**, 54423 (2011).

[78] G. A. M. Alexander G. Gurevich, *Magnetization Oscillations and Waves*, 1st Editio (CRC Press, 1996).

[79] V. Haspot, P. Noël, J.-P. Attané, L. Vila, M. Bibes, A. Anane, and A. Barthélémy, Temperature dependence of the Gilbert damping of ${\mathrm{La}}_{0.7}{\mathrm{Sr}}_{0.3}{\mathrm{MnO}}_{3}$ thin films, Phys. Rev. Mater. **6**, 24406 (2022).

[80] C. Bell, S. Milikisyants, M. Huber, and J. Aarts, Spin dynamics in a superconductor-ferromagnet proximity system, Phys. Rev. Lett. **100**, 1 (2008).

[81] H. Behner, K. Rührnschopf, G. Wedler, and W. Rauch, Surface reactions and long time stability of YBCO thin films, Physica C Supercond. **208**, 419 (1993).

[82] Y. Wang et al., Thickness and temperature-dependent damping in La0.67Sr0.33MnO3 epitaxial films, Appl. Phys. Lett. **123**, (2023).
[83] Y. Tanaka and S. Kashiwaya, Local density of states of quasiparticles near the interface of nonuniform d-wave superconductors., Phys. Rev. B Condens. Matter **53**, 9371 (1996).
[84] S. Kashiwaya and Y. Tanaka, Theory for tunneling spectroscopy of anisotropic superconductors, Phys. Rev. B Condens. Matter Mater. Phys. **53**, 2667 (1996).
[85] S. Kashiwaya and Y. Tanaka, Tunnelling effects on surface bound states in unconventional superconductors, Reports on Progress in Physics **63**, 1641 (2000).
[86] M. A. Silaev, Large enhancement of spin pumping due to the surface bound states in normal metal/superconductor structures, Phys. Rev. B **102**, 180502 (2020).
[87] V. Prokić, A. I. Buzdin, and L. Dobrosavljević-Grujić, Theory of the \ensuremath{\pi} junctions formed in atomic-scale superconductor/ferromagnet superlattices, Phys. Rev. B **59**, 587 (1999).
[88] L. N. Bulaevskii and M. V Zyskin, Energy gap in layered superconductors, Phys. Rev. B **42**, 10230 (1990).
[89] A. I. Buzdin, V. P. Damjanović, and A. Yu. Simonov, Thermodynamic properties of atomic superconductor--normal-metal multilayers, Phys. Rev. B **45**, 7499 (1992).
[90] D. Yu. Gusakova, A. A. Golubov, M. Yu. Kupriyanov, and A. Buzdin, Density of states in SF bilayers with arbitrary strength of magnetic scattering, Journal of Experimental and Theoretical Physics Letters **83**, 327 (2006).